\documentclass[]{emulateapj}

\newcommand{\myemail}{t.hayashi@nao.ac.jp}
\shorttitle{Multi-freq.~polarimetric Imaging of Radio-loud BAL Quasars}

\begin{document}
\title{VLBA Multi-frequency Polarimetric imaging\\ of Radio-loud Broad Absorption Line Quasars}
\author{Takayuki J. Hayashi\altaffilmark{1,2}, Akihiro Doi\altaffilmark{3,4}, and Hiroshi Nagai\altaffilmark{1}}
\email{\myemail}

\altaffiltext{1}{National Astronomical Observatory of Japan, 2-21-1 Osawa, Mitaka, Tokyo, 181-8588, Japan}
\altaffiltext{2}{Department of Astronomy, The University of Tokyo, 7-3-1 Hongo, Bunkyo, Tokyo, 133-0033, Japan}
\altaffiltext{3}{Institute of Space and Astronautical Science, Japan
Aerospace Exploration Agency, 3-1-1 Yoshinodai, Chuo, Sagamihara, Kanagawa, 252-5210, Japan}
\altaffiltext{4}{Department of Space and Astronautical Science, The Graduate University for Advanced Studies, 3-1-1 Yoshinodai,  Chuo, Sagamihara, Kanagawa, 229-8510, Japan}

\begin{abstract}
We conducted the first multi-frequency polarimetric imaging of four broad absorption line (BAL) quasars using Very Long Baseline Array at milli-arcsecond resolutions to investigate the inclination of the non-thermal jet and test the hypothesis that radio sources in BAL quasars are still young.
Among these four sources, J0928+446, J1018+0530, and J1405+4056 show one-sided structures in parsec scales, and polarized emission detected in the core.
These characteristics are consistent with those of blazars.
We set constraints on viewing angles to $<$66~deg for these jets, in the framework of a Doppler beaming effect.
J1159+0112 exhibits an unpolarized gigahertz peaked spectrum component and several discrete blobs with steep spectra on  both sides of the central component across $\sim$1~kpc. 
These properties are consistent with those of young radio sources.
We discuss the structures of jets and AGN wind. 
\end{abstract}


\section{INTRODUCTION}\label{sec:intro}

Quasars showing broad absorption troughs of resonance lines (e.g., \ion{Mg}{2}, \ion{C}{4}) in their rest ultra violet spectra are called broad absorption line (BAL) quasars \citep[e.g.,][]{1991ApJ...373...23W}. 
BAL quasars are classified into two categories, HiBAL and LoBAL quasars, depending on whether they show both high-ionization and low-ionization troughs or only low-ionization troughs. 
Ionized wind from an accretion disk of an active galactic nucleus (AGN) is the most plausible candidate for the origin of BAL to account for its high velocities (several thousand km s$^{-1}$ to $\sim$0.1$c$ typically). 
For the quasar sample from the  Sloan Digital Sky Survey Third Data Release (SDSS DR3; \citealt{2005AJ....129.1755A}), the fraction of BAL quasars is 26\% \citep{2006ApJS..165....1T}, 
which is a key to understanding the origin of BAL quasars.
To explain the fraction, two scenarios have been proposed: an orientation scheme and an evolution scheme.

In the orientation scheme, all quasars possess the AGN wind and whether BAL is observed or not can be attributed to the viewing angles to the sources. 
The AGN wind is produced by radiation pressure nearly parallel to the disk \citep{2000ApJ...543..686P,2000ApJ...545...63E}, based on support in spectropolarimetry at rest ultra violet \citep{1995ApJ...448L..73G,1995ApJ...448L..77C}.
In the model, BAL troughs can only be seen in quasars whose edge-on disk wind points towards the observer.
In contrast, \cite{2006ApJ...639..716Z} found a number of radio-detected BAL quasars showing rapid radio-flux variation which indicates a Doppler beaming effect on pole-on viewed jets \citep{1979ApJ...232...34B,1995PASP..107..803U}.
Moreover, widely distributed radio spectral indices \citep{2000ApJ...538...72B,2008MNRAS.388.1853M,2009PASJ...61.1389D,2011MNRAS.412..213F,2011ApJ...743...71D,2012A&A...542A..13B} also suggest that at least a portion of BAL quasars are blazar-type objects with flat-spectrum radio cores.  
The simple orientation scheme might not explain all BAL quasars.

The evolution scheme ascribes the ratio of BAL to non-BAL quasars to the duration of time when quasars possess the AGN wind \citep{2000ApJ...538...72B}.
Most BAL quasars detected by the Faint Images of the Radio Sky at Twenty centimeters survey (FIRST survey; \citealt{1995ApJ...450..559B}) are point-like sources \citep{2000ApJ...538...72B}.
The number of Fanaroff-Riley type-II (FR II; \citealt{1974MNRAS.167P..31F}) radio sources in BAL quasars is roughly 10 times less common than that in all SDSS quasars \citep{2006ApJ...641..210G}.
In addition,  \cite{2008MNRAS.388.1853M} reported that a significant fraction of BAL quasars show radio spectra as found in gigahertz-peaked spectrum (GPS) or compact steep spectrum (CSS) radio sources that are candidates for young radio sources \citep{2000MNRAS.319..445S,2002NewAR..46..263C,2006ApJ...648..148N}.
Quasars might show their BAL when they are young before they have large-scale jets.
Thus, radio observations to measure the age of a source is an important approach for understanding BAL quasars in terms of the evolution scheme. 

In contrast to the above argument, GPS/CSS sources as young radio sources are sometimes indistinguishable from blazars by radio observations with arcsecond scale resolution.
Usually, blazars have a flat or inverted spectrum up to high frequencies, while young radio sources have an optically thin steep spectrum in the gigahertz regime.
However, blazars may show a convex spectrum similar to the GPS spectrum during a flare \citep{2005A&A...435..839T,2007A&A...469..451T}.
Thus, observations with limited frequency coverage cannot distinguish GPS/CSS sources from blazars.
Even in these situations, blazars and young radio sources still display characteristics different from each other \citep{2005A&A...432...31T,2007A&A...475..813O}.
Blazars appear as one-sided structures in milli-arcsecond (mas) resolutions, long-term variability \citep{2005A&A...435..839T,2007A&A...469..451T,2007A&A...469..899H,2008A&A...485...51H}, and a high degree of polarization.
On the other hand, young radio sources have lobes with sub-relativistic speeds, which appear as two-sided structures in mas resolution, little variability, and little polarization.
To distinguish blazars and young radio sources, high-resolution direct imaging is important.
Then, very long baseline interferometry (VLBI) is efficient to test the evolution scheme.
In addition, VLBI is also one of the most direct ways to test the orientation scheme because radio structures on a mas scale  include the information about the viewing angle.

There have been several studies of BAL quasars using VLBI \citep{2003A&A...397L..13J, 2008evn..confE..19M, 2008MNRAS.391..246L, 2009PASJ...61.1389D, 2009ApJ...706..851R, 2010ApJ...718.1345K, 2010evn..confE..36B,2011arXiv1106.5916G, 2012MNRAS.419L..74Y}.
Nonthermal jets and the AGN wind coexist simultaneously at least in radio-loud BAL quasars.
Furthermore, there are some inverted-spectrum sources, which are interpreted as young radio sources or Doppler-beamed sources having pole-on-viewed relativistic jet \citep{2009PASJ...61.1389D}.
Both one-sided and two-sided jet structures have been found \citep{2003A&A...397L..13J,2008evn..confE..19M}.
Only one polarization study has been conducted by \cite{2008MNRAS.391..246L}, which reported no difference in radio morphologies and polarization features between flat- and steep-spectrum sources.
Two sources were observed at more than three frequencies and show signatures of interaction with the interstellar medium (ISM). \cite{2010ApJ...718.1345K} reported disturbed morphology of J1048+3457.
\cite{2009ApJ...706..851R} discussed  the ram pressure of the external medium on the Mrk~231 radio source. 
Thus far, general mas-scale radio properties of BAL quasars are still unclear from the point of view of the orientation scheme and the evolution scheme.  

In the present paper, we report the result of multi-frequency polarimetric imaging observations using Very Long Baseline Array (VLBA) for four radio-loud BAL quasars with flat or inverted spectra on sub-arcsecond scale.
We describe our sample sources in Section~2.
The observation and data reduction are described in Sections~3 and 4, respectively.
The results are presented in Section~5. 
We discuss the structures of parsec-scale radio jets and the AGN wind, the orientation scheme, and the evolution scheme in Section~6.
Finally, our conclusions are summarized in Section~7.  
Throughout this paper, we adopt a cosmology consistent with WMAP results of $h=0.71$,  $\Omega _M=0.27$, and $\Omega_\Lambda=0.75$ \citep{2003ApJS..148..175S}.
The angular scale of 1~mas corresponds to 8.4~parsec (pc) at the distances of our targets at $z\sim2$.

\section{TARGET SOURCES}\label{sec:source}

We selected a sample for the VLBA observation from 20~sources detected in the first systematic VLBI observation at 8.4~GHz \citep{2009PASJ...61.1389D} using the Optically ConnecTed Array for VLBI Exploration project (OCTAVE; \citealt{2008evn..confE..41K}) operated by the Japanese VLBI Network (JVN;  \citealt{2008evn..confE..75F}).
The OCTAVE sample consisted of SDSS-DR3 BAL quasars in \cite{2006ApJS..165....1T}, which hosted radio counterparts in the FIRST survey with peak flux densities of more than 100~mJy. 
Then, we selected the target sources that have (i) flux density of more than 100 mJy in the OCTAVE observation, (ii) expected polarized flux density\footnote{
Expected polarized flux density is defined as a product of degree of polarization provided by the NRAO VLA Sky Survey at 1.4~GHz (NVSS; \citealt{1998AJ....115.1693C}) and flux density obtained by the OCTAVE observation at 8.4~GHz.
}
of more than 1 mJy, and (iii) flat or inverted spectra\footnote{
Throughout this paper, spectral index, $\alpha$, is defied as $\alpha = \Delta \ln S_\nu/\Delta \ln \nu$, where $S_\nu$ is the flux density at the frequency, $\nu$. 
} ($\alpha > -0.5$) in \cite{2009PASJ...61.1389D}.
The sample is listed in Table~\ref{tbl:sample}.

\section{OBSERVATION}\label{sec:obs}

The multi-frequency polarimetric imaging was conducted at 1.7-, 5-, and 8-GHz bands using VLBA on 2010 June 25 (project code BD137).
The observation was carried out over 10~hours.  Each source was observed at the three bands with 6--10~minutes scan at 3--4 different hour angles.
This leads to the quasi-similar $u$-$v$ coverages at each band.
An aggregate bit rate of 128~Mbps was used; each band consisted of two 8-MHz wide, full polarization intermediate frequencies (IFs) centered at 1.663 and 1.671~GHz at 1.7-GHz band, 4.644 and 5.095~GHz at  5-GHz  band, and 8.111 and 8.580~GHz at  8-GHz  band.
We integrated two IFs in each band to make Stokes $I$ maps, while we produced the polarization map of each IF separately.

It is important to select a suitable setting for the IFs to determine $n$-$\pi$ ambiguity ($n=0, \pm 1, \pm 2 \cdots$) in polarization angle and to measure  Faraday rotation measure (RM).
The upper limit of measurable RM is determined by $\pi/2> (1+z)^2 |{\rm RM_{obs}}| \Delta_{\rm m} \lambda ^2$, where ${\rm RM_{obs}}$ and $\Delta_{\rm m}\lambda^2$ are the RM at the observer frame and the minimum separation of the square of observing wavelength, respectively.
Then, we obtain the maximum measurable RM as $\sim$10,000~rad~m$^{-2}$ in the case of our setting.
The maximum value at rest frame becomes $\sim$90,000~rad~m$^{-2}$ for our sample sources at $z\sim 2$.

\section{DATA REDUCTION}\label{sec:data}

\subsection{{\it A priori} Calibration and Imaging Process }

Data reduction was performed with a standard procedure using the Astronomical Image Processing System ({\tt AIPS}; \citealt{2003ASSL..285..109G}) software developed at the National Radio Astronomy Observatory (NRAO).
Amplitude calibration was performed using the measurements of system noise temperature during the observation and gains provided by each station.
We also corrected the amplitude attenuation due to atmospheric opacity.
Fringe fitting was performed after the Earth orientation parameters and ionospheric dispersive delay were corrected.
Finally, bandpass calibration for both amplitude and phase was performed.
All sources were detected on all baselines at all frequencies except for the 1.7-GHz band at Hancock station where system temperature was not obtained properly due to radio frequency interference.

Imaging processes were performed using the {\tt difmap} software \citep{1997ASPC..125...77S}.
We conducted self-calibration to derive the antenna-based amplitude corrections. 
The {\tt difmap} does not solve for gain time variation for RR and LL visibilities separately.
Hence, we constructed RR or LL model by the {\tt difmap} and then performed self-calibration for Stokes $I$ by the {\tt AIPS}, which corrects gain time variation for RR and LL visibilities separately \citep{EVNmemo78}.
The error on the absolute flux density scale was generally $\sim$5\%.

\subsection{Calibration of Polarization}

We corrected  RL and LR delay offsets using the bright polarized source, J0854+2006 (OJ~287), and corrected instrumental polarization ($D$-term) using the compact unpolarized source, J1407+2827 (OQ~208).
After $D$-term was calibrated, we confirmed that OQ~208 had almost  become unpolarized.
We estimated the error on the absolute flux density scale for polarization was within 10\% including residual $D$-term.

\begin{figure}
\includegraphics[width=1.0\textwidth]{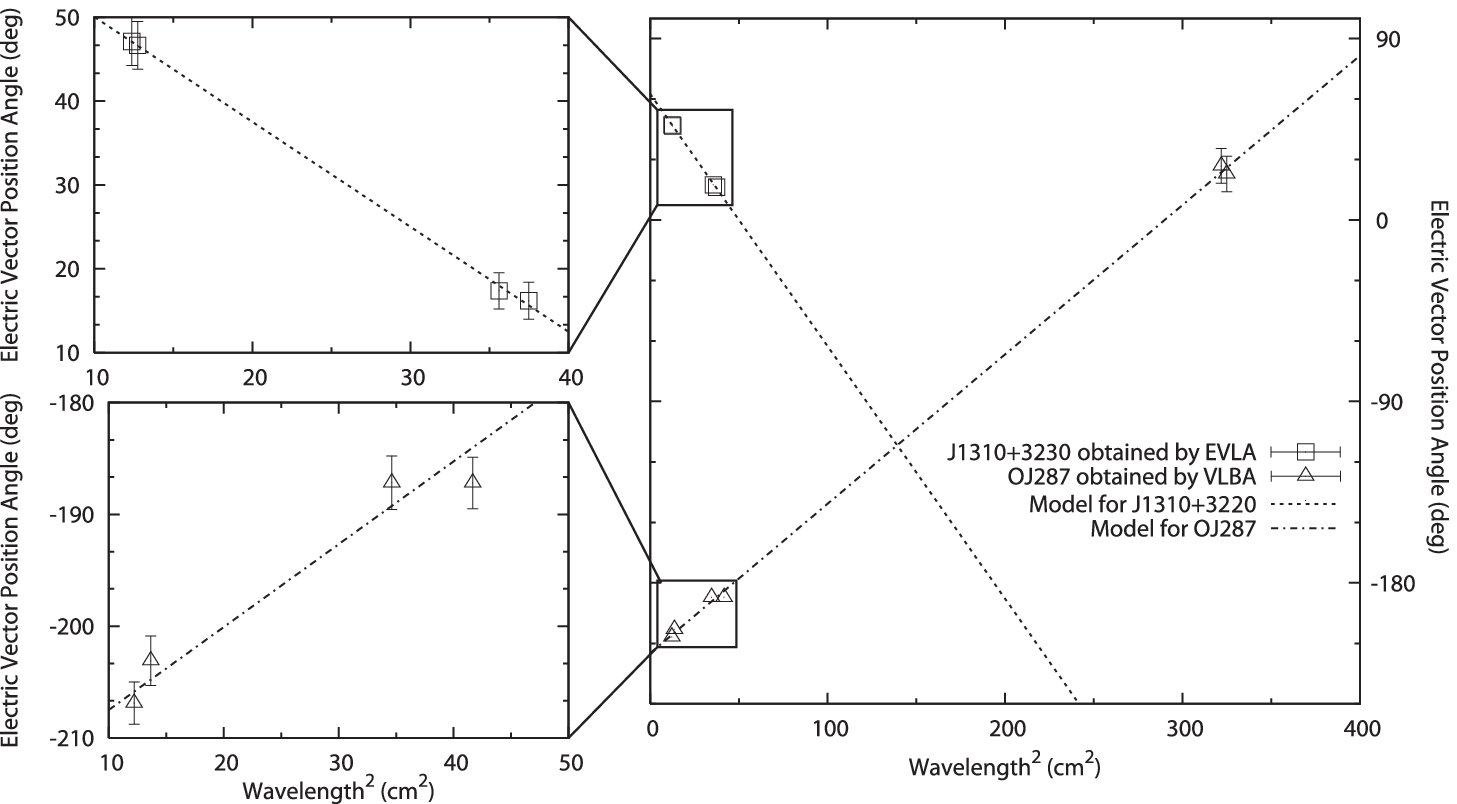}
\figcaption{
	EVPA for polarization calibrator sources of J1310+3220 obtained by EVLA observation and OJ~287 obtained by our VLBA observation.
    We refer to EVPA of J1310+3220 (open squares) for that of OJ~287 (open triangles) and target sources.
	Reference value at 1.7~GHz is obtained by extrapolation of the value of 5 and 8~GHz (dotted line).
	EVPA of OJ~287 at 1.7~GHz is in excellent agreement with EVPA of OJ~287 extrapolated from 5 and 8~GHz (dot-dashed line).
\label{fig:EVPAcal}
}
\end{figure}

An unknown phase offset between L and R polarizations for a reference antenna was corrected using the observed electric vector position angle (EVPA) of J1310+3220 which was observed by EVLA at 5- and 8-GHz bands on 2010 July 15 and June 22, respectively, under the project named TPOL.
Each band consists of two 128-MHz wide IFs centered at 4.896 and 5.024~GHz at 5-GHz band, 8.395 and 8.523~GHz at 8-GHz band.
Using the data obtained by UMRAO\footnote{
The data was obtained by the University of Michigan Radio Astronomy Observatory (UMRAO), which was kindly provided by M.~Aller.
}, 
we confirmed that during the months of June and July there was no significant variability in their total flux densities, degree of polarization, and EVPA.
In addition, total and polarized flux density of J1310+3220 obtained by VLBA were similar to that obtained by EVLA.
We derived the integrated EVPA for J1310+3220 at 1.7-GHz band by extrapolating from EVPA at 5- and 8-GHz bands because EVPA would be affected by RM.
After the EVPA correction,
EVPA of the other EVPA calibrator, OJ~287, at 1.7-GHz band  was consistent with that extrapolated from EVPA at 5- and 8-GHz bands obtained using VLBA (Figure~\ref{fig:EVPAcal}).
Hence, our EVPA calibrations seemed to be performed well.  

The errors of EVPA are the root sum square of flux density measurement errors and fitting error to derive RM for the calibrator source.
Because polarimetry at low frequency is affected by ionospheric Faraday rotation, we checked the total electron content of the ionosphere during our observation;
the typical variation of ionospheric RM within a scan was $|{\rm RM_{obs}}|< 0.5$~rad~m$^{-2}$ and between scans was $|{\rm RM_{obs}}|< 3$~rad~m$^{-2}$.
These values are equivalent to $\Delta {\rm EVPA}< 0.02$~rad and $\Delta {\rm EVPA}< 0.12$~rad at a wavelength of 20~cm, which do not affect the estimation of the EVPA significantly even at 1.7-GHz band.

\section{RESULTS}\label{sec:result}

\subsection{Morphology}\label{sec:Imap}
\begin{figure}
\includegraphics[width=0.5\textwidth]{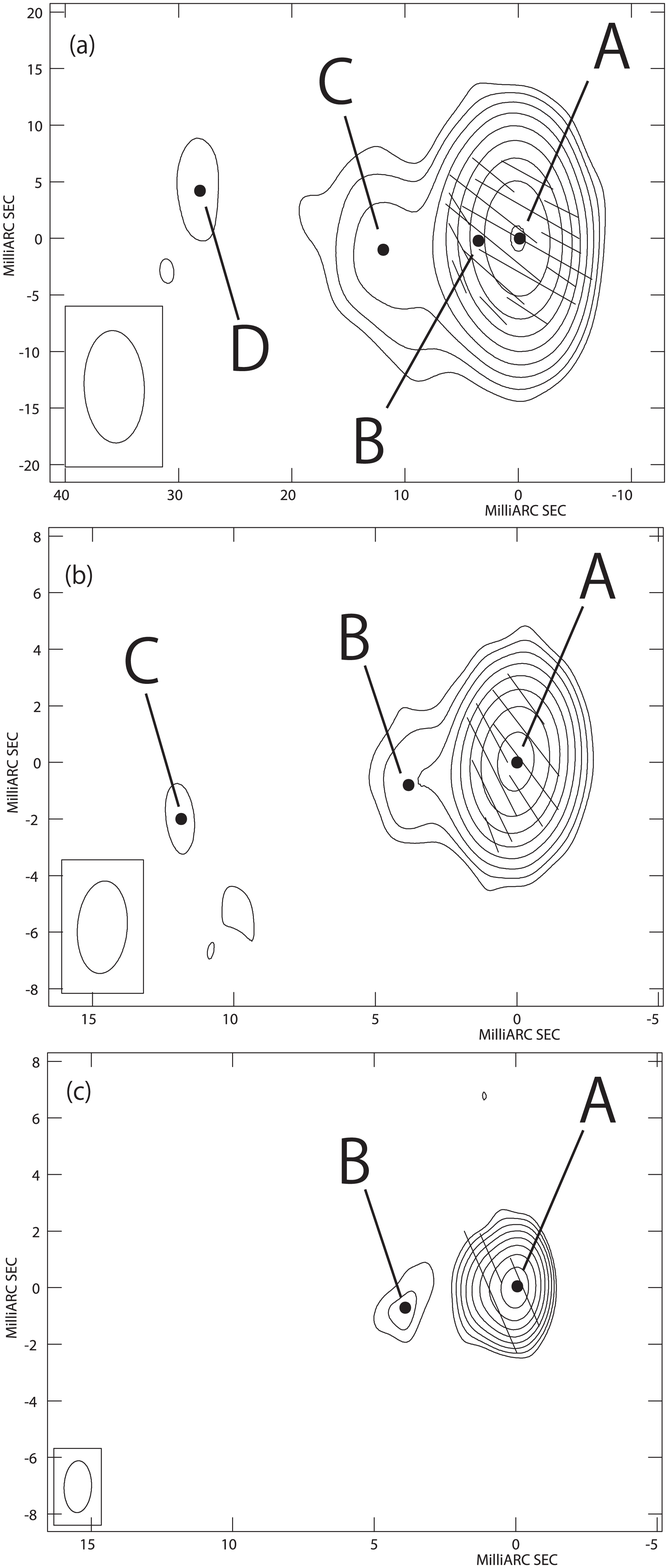}
\figcaption{\footnotesize
VLBA images of J0928+4446 observed on 2010 June 25 with superposed polarization vectors.
Vector lengths are proportional to the polarized flux density (1~mas corresponds to 0.5~mJy~beam$^{-1}$)
	(a) The 1.663~GHz Stokes $I$ map. The restoring beam is $9.9 \times 5.3$~mas at PA$=2.5^\circ$, 
	the contour levels are $(1, 2, 4, 8, \cdots) \times 0.63$~mJy beam$^{-1}$, and the peak flux density is 168~mJy beam$^{-1}$. 
	(b) The 4.644~GHz Stokes $I$ map. The restoring beam is $3.3 \times 1.8$~mas at PA$=2.8^\circ$, 
	the contour levels are $(1, 2, 4, 8, \cdots) \times 1.1$~mJy beam$^{-1}$, and the peak flux density is 179~mJy beam$^{-1}$. 
	(c) The 8.111~GHz Stokes $I$ map. The restoring beam is $1.8 \times 0.97$~mas at PA$=-2.3^\circ$, 
	the contour levels are $(1, 2, 4, 8, \cdots) \times 1.0$~mJy beam$^{-1}$, and the peak flux density is 196~mJy beam$^{-1}$. 
\label{fig:J0928_I}
}
\end{figure}

\begin{figure}
\includegraphics[width=1.0\textwidth]{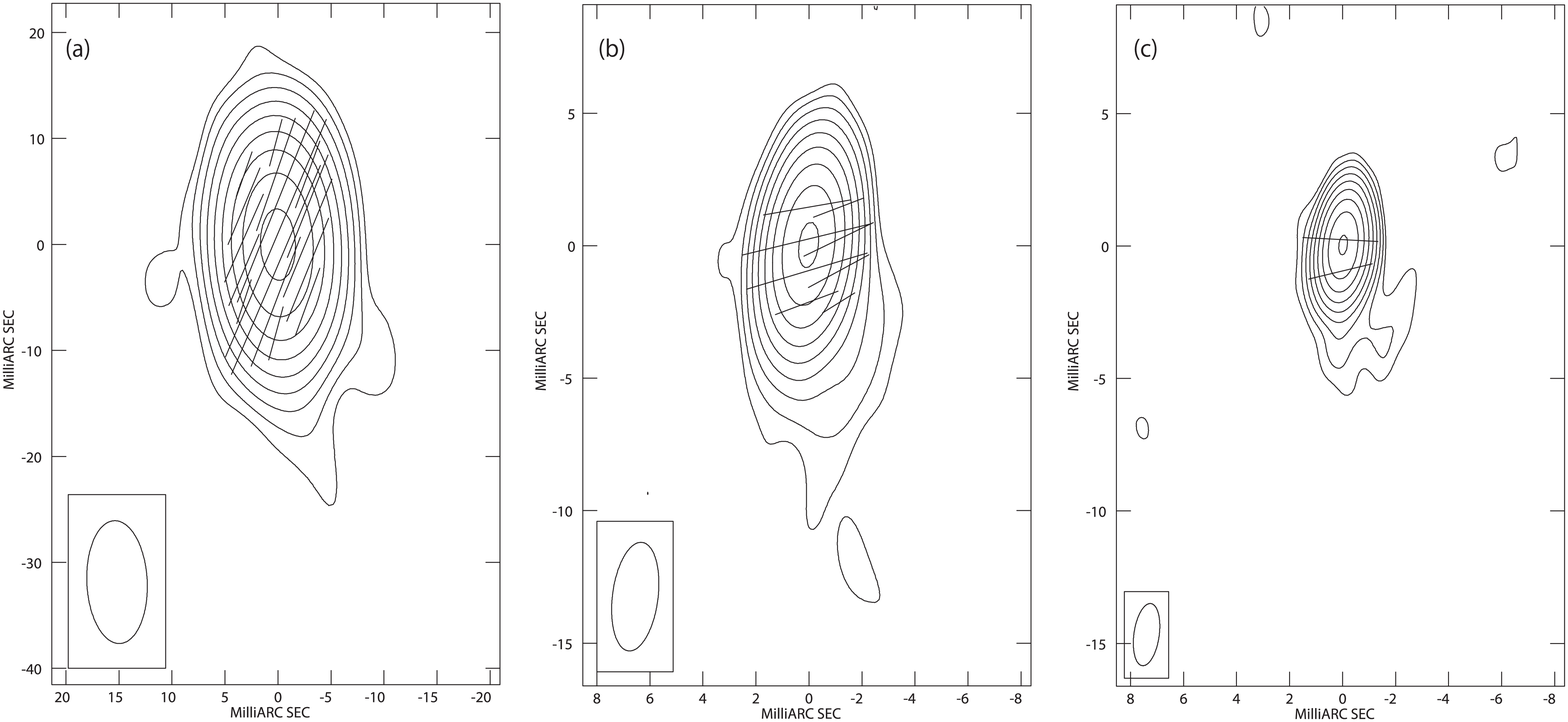}
\figcaption{\footnotesize
VLBA images of J1018+0530 observed on 2010 June 25 with superposed polarization vectors.
Vector lengths are proportional to the polarized flux density (1~mas corresponds to 0.5~mJy~beam$^{-1}$)
	(a) The 1.663~GHz Stokes $I$ map. The restoring beam is $12 \times 5.7$~mas at PA$=2.7^\circ$, 
	the contour levels are $(1, 2, 4, 8, \cdots) \times 1.1$~mJy beam$^{-1}$, and the peak flux density is 344~mJy beam$^{-1}$. 
	(b) The 4.644~GHz Stokes $I$ map. The restoring beam is $4.1 \times 1.7$~mas at PA$=-7.2^\circ$, 
	the contour levels are $(1, 2, 4, 8, \cdots) \times 1.2$~mJy beam$^{-1}$, and the peak flux density is 334~mJy beam$^{-1}$. 
	(c) The 8.111~GHz Stokes $I$ map. The restoring beam is $2.4 \times .0.95$~mas at PA$=-8.0^\circ$, 
	the contour levels are $(1, 2, 4, 8, \cdots) \times 1.2$~mJy beam$^{-1}$, and the peak flux density is 339~mJy beam$^{-1}$. 
\label{fig:J1018_I}
}
\end{figure}

\begin{figure}
\includegraphics[width=1.0\textwidth]{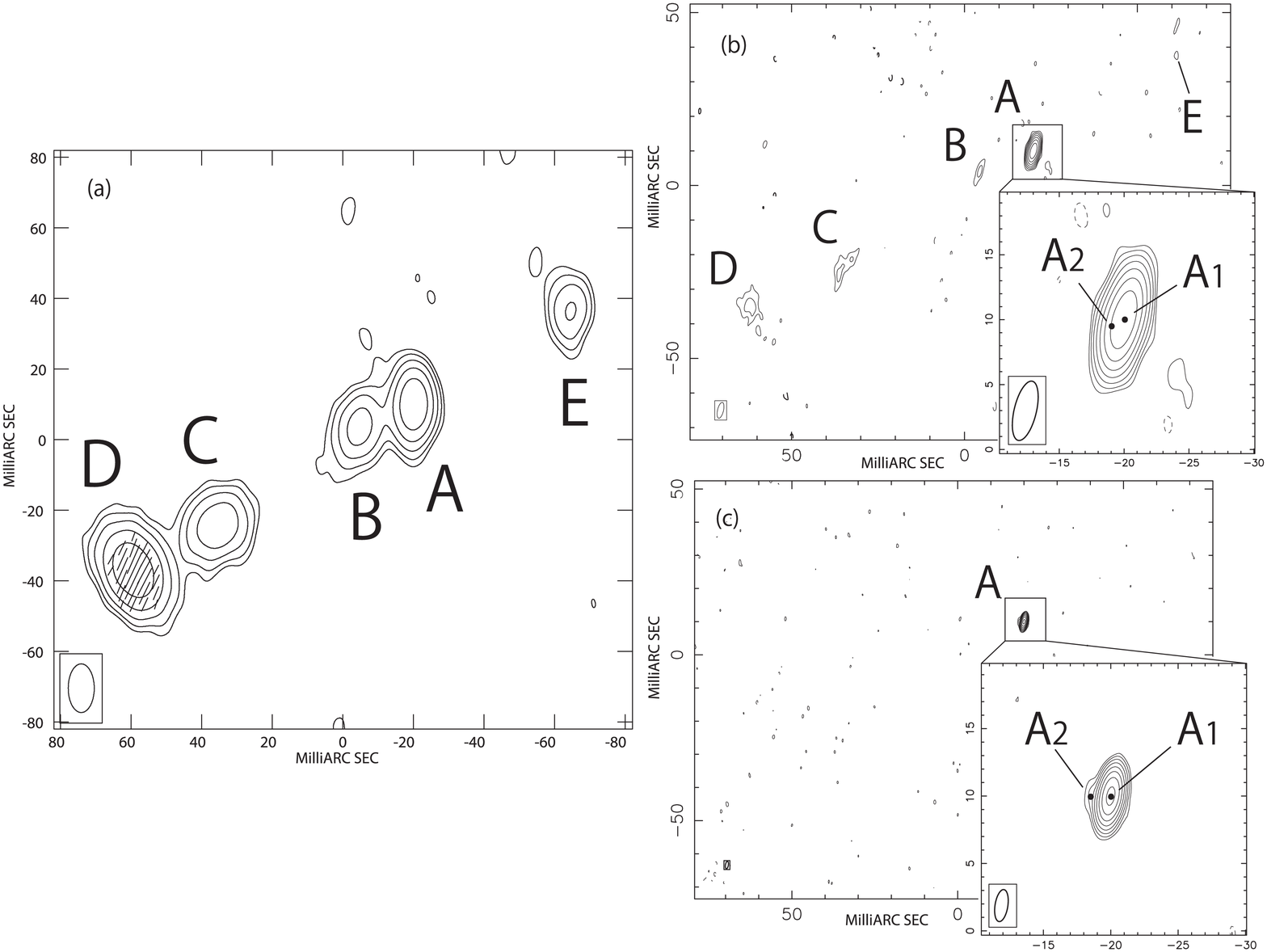}
\figcaption{\footnotesize
VLBA images of J1159+0112 observed on 2010 June 25 with superposed polarization vectors.
Vector lengths are proportional to the polarized flux density (1~mas corresponds to 0.5~mJy~beam$^{-1}$)
	(a) The 1.663~GHz Stokes $I$ map. The restoring beam is $14 \times 7.4$~mas at PA$=0.080^\circ$, 
	the contour levels are $(1, 2, 4, 8, \cdots) \times 1.3$~mJy beam$^{-1}$, and the peak flux density is 36.4~mJy beam$^{-1}$. 
	(b) The 4.9~GHz image. The restoring beam is $4.7 \times 1.7$~mas at PA$=-14^\circ$,
	the contour levels are $(1, 2, 4, 8, \cdots) \times 0.78$~mJy beam$^{-1}$, and the peak flux density is 93.0~mJy beam$^{-1}$.
	(c) The 8.3~GHz image. The restoring beam is $2.4 \times 0.95$~mas at PA$=-8.9^\circ$,
	the contour levels are $(1, 2, 4, 8, \cdots) \times 0.76$~mJy beam$^{-1}$, and the peak flux density is 122~mJy beam$^{-1}$.
\label{fig:J1159_I}
}
\end{figure}

\begin{figure}
\includegraphics[width=0.5\textwidth]{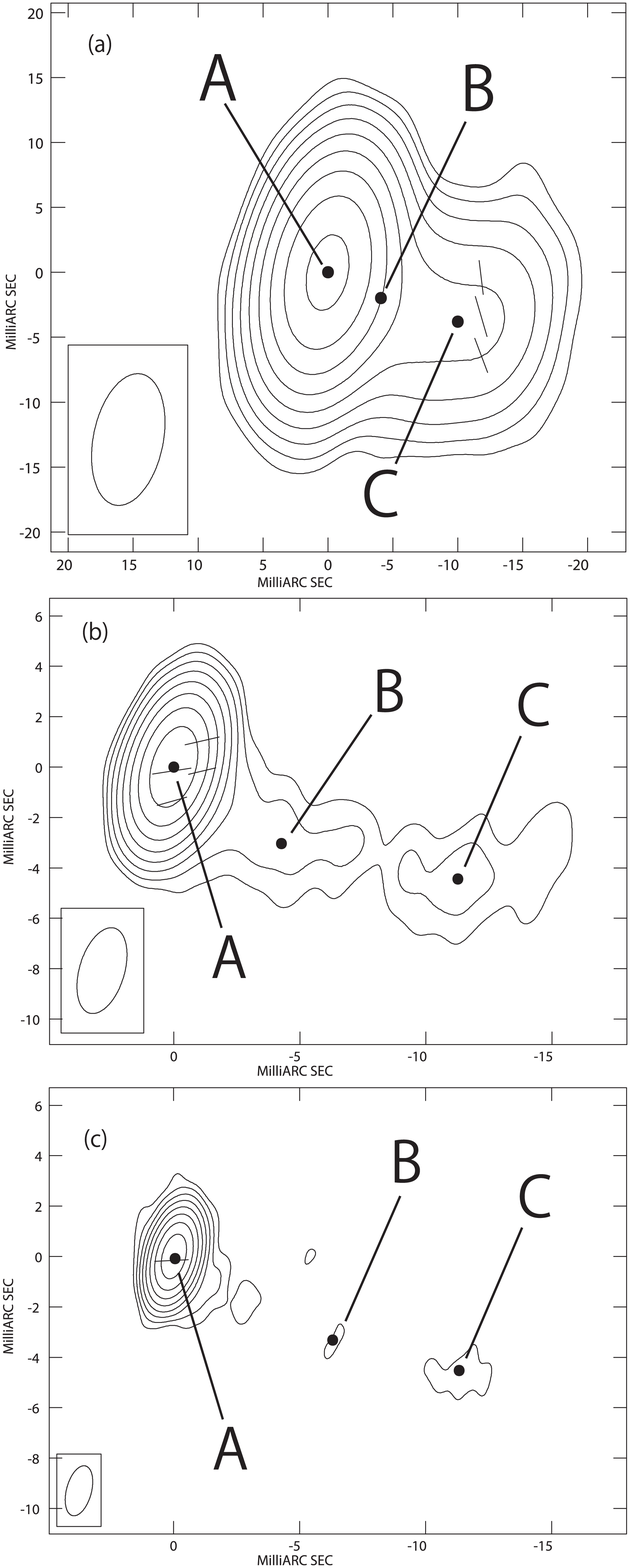}
\figcaption{\footnotesize
VLBA images of J1405+4056 observed on 2010 June 25 with superposed polarization vectors.
Vector lengths are proportional to the polarized flux density (1~mas corresponds to 0.5~mJy~beam$^{-1}$).
	(a) The 1.663~GHz Stokes $I$ map. The restoring beam is $10 \times 5.4$~mas at PA$=-11^\circ$, 
	the contour levels are $(1, 2, 4, 8, \cdots) \times 0.69$~mJy beam$^{-1}$, and the peak flux density is 220~mJy beam$^{-1}$. 
	(b) The 4.644~GHz Stokes $I$ map. The restoring beam is $3.5 \times 1.8$~mas at PA$=-17^\circ$, 
	the contour levels are $(1, 2, 4, 8, \cdots) \times 0.94$~mJy beam$^{-1}$, and the peak flux density is 220~mJy beam$^{-1}$. 
	(c) The 8.111~GHz Stokes $I$ map. The restoring beam is $2.0 \times 1.0$~mas at PA$=-15^\circ$, 
	the contour levels are $(1, 2, 4, 8, \cdots) \times 0.89$~mJy beam$^{-1}$, and the peak flux density is 192~mJy beam$^{-1}$. 
\label{fig:J1405_I}
}
\end{figure}

Stokes $I$ maps of the target sources at 1.663, 4.644, and 8.111~GHz are shown in Figures~\ref{fig:J0928_I}--\ref{fig:J1405_I}.
Flux densities of each component were measured by fitting with a Gaussian model profile and the spectral indices were calculated for each component (Table~\ref{tbl:flux}). 
 
{\bf J0928+4446}: 
The radio structure in Figure~\ref{fig:J0928_I} is consistent with that obtained by the VLBA Imaging and Polarimetry Survey (VIPS; \citealt{2007ApJ...658..203H}) at 5~GHz.
The spectral index of each component indicates (see Table~\ref{tbl:flux}) an inverted-spectrum core (the component A) and steep spectrum one-side jets (the components B-D).
Its morphology can be classified as a one-sided structure.

{\bf J1018+0530}: 
The images at 5 and 8~GHz show an extended emission, which emerges at position angle (PA) of $\sim$$-170$~deg (Figure~\ref{fig:J1018_I}), while the source is unresolved at 1.7~GHz.
The pc-scale radio structure is dominated by an inverted-spectrum core (Table~\ref{tbl:flux}). 
The extended emission found at 5 and 8~GHz is a jet.
Its morphology can be classified as a one-sided structure.

{\bf J1159+0112}: 
Table~\ref{tbl:flux} indicates inverted spectrum at the component A, which is a radio core.
Additionally, images at 1.7 and 5~GHz show the linear alignment of several discrete components that extend $\sim$90~mas towards the southeast and a significant counter feature $\sim$50~mas northwest across $\sim$1~kpc in total (Figure~\ref{fig:J1159_I}). 
The southeast components are consistent with the previous study by \cite{2008evn..confE..19M}.
They show steep spectra (Table~\ref{tbl:flux}) and thus morphology that can be classified as a two-sided structure.
Although \cite{2008evn..confE..19M} also reported two symmetrical extensions close to the core located towards the northwest and the southeast, no such structure was detected by our observation.
Instead, we found the components A1 and A2 both at 5 and 8~GHz.

\begin{figure}
	\includegraphics[width=0.5\textwidth]{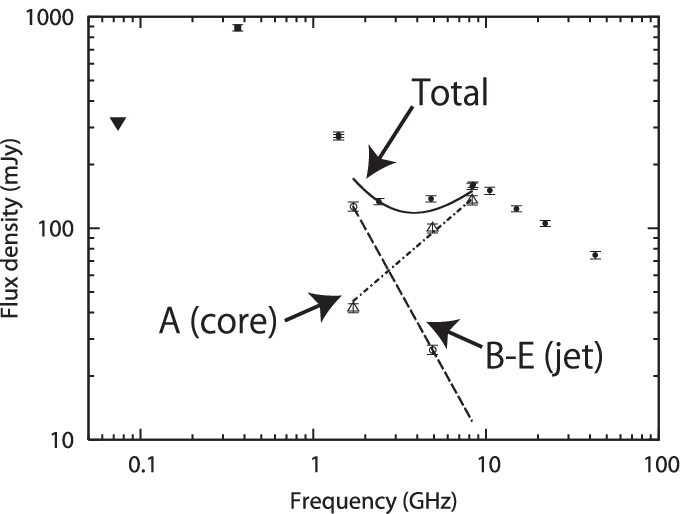}
	\figcaption{
	Radio spectrum of J1159+0112.
	Observations using VLA (filled circle; \citealt{1998AJ....115.1693C,1995ApJ...450..559B,2008MNRAS.388.1853M}),
	flux density from the Texas survey (filled square at 365~MHz; \cite{1996AJ....111.1945D}), and our observation using VLBA are shown.
	A triangle means 3$\sigma$ upper limit at 74~MHz \citep{2007AJ....134.1245C}.
	Flux density obtained by our observation is decomposed into that of the radio core (open triangle; the component A) and jets (open circle; the components B--E).
	Power law fit to the core flux density (dot-dashed), jet flux density (dashed), and sum of them (thick) are also illustrated.
	Although our observation was able to set 3$\sigma$ upper limit of 0.8~mJy for jet components at 8.3~GHz, it might be underestimated because the source could be resolved out.
	\label{fig:SED_J1159}
}
\end{figure}

The radio spectrum of J1159+0112 (Figure~\ref{fig:SED_J1159}) can be represented by double-peaked spectra, peaked at a few hundred~MHz and $\sim$10~GHz: a steep spectrum and an inverted spectrum in the range of our VLBA observations. 
The GHz-peaked component originates in the radio core (the component A), while the steep spectrum components originate in the extended structure with several discrete blobs (the components B--E).

{\bf J1405+4056}: 
The structure in Figures~\ref{fig:J1405_I}  is consistent with that obtained by the VIPS.
Table~\ref{tbl:flux} indicates that the radio structure consists of an inverted-spectrum core (the component A) and steep spectrum one-side jets (the components B and C).
Its morphology can be classified as a one-sided structure.

\subsection{Polarization}\label{sec:pol}

Figures~\ref{fig:J0928_I}--\ref{fig:J1405_I} show polarization vectors overlaid on the Stokes $I$ maps.
Polarized flux densities and degree of polarization at each component averaged within a band is listed in Table~\ref{tbl:pol_flux}.
Errors are the root sum square of calibration uncertainties of 10\% and fitting error in the {\tt AIPS} task {\tt IMFIT}.
EVPA at components whose polarized flux densities are detected is shown in Table~\ref{tbl:EVPA}.

Point-like polarized emissions in the radio core were detected for all the sources except for J1159+0112.
Polarization of J1159+0112 was detected in the component D with a very high degree of polarization of $11.4\pm 1.7$\% at 1.7~GHz.
This will be discussed in Section~\ref{sec:discussionJ1159+0112}.
For J1405+4056, no polarized emission at 1.663 and 1.671~GHz was detected at the core.
Polarized emission at low frequencies could suffer from depolarization within a bandwidth and/or within a beamwidth.
To make the source depolarized within a 8-MHz bandwidth at 1.7~GHz, RM more than 10,000~rad~m$^{-2}$ at observer-frame is needed.
Such a high RM is reported in a BAL quasar, J1624+3758 \citep{2012A&A...542A..13B}.
Alternatively, even if RM is less than 10,000~rad~m$^{-2}$, inhomogeneous spatial distribution of magnetic field and electron density could cause disordered EVPA distribution across the beam.
Then, lower-frequency polarized emissions tend to be smeared out because of larger beam sizes.  
On the jet component of J1405+4056, we found a hint of polarized flux density only at 1.663 GHz (Figure~\ref{fig:J1405_I}) but the same polarization structure was not detected at the other frequencies including 1.671~GHz.

\subsection{Faraday Rotation Measure}\label{sec:RM}

\begin{figure}
\includegraphics[width=1.0\textwidth]{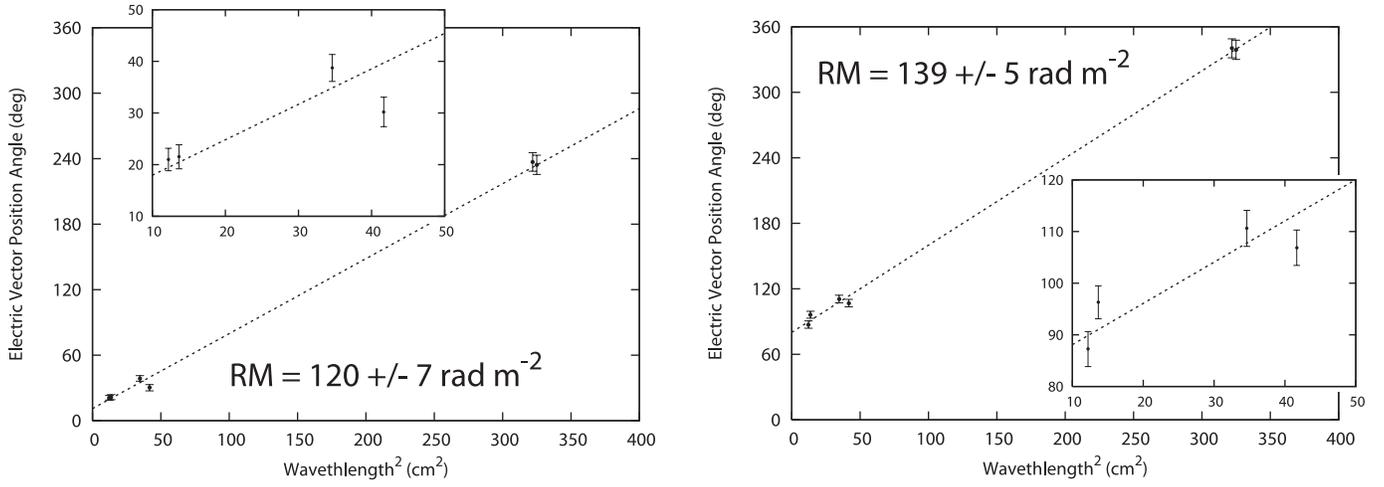}
\figcaption{
Fits of rotation measure for J0928+4446 (left) and J1018+0530 (right).
Because polarization structures of the sources were simple, we derived RM using EVPA obtained by integrated flux densities of Stokes $Q$ and $U$ maps.
\label{fig:RM}
}
\end{figure}

RM denotes the dependence of EVPA on wavelength. 
The result of RM fits is shown in Figure~\ref{fig:RM} and Table~\ref{tbl:EVPA}.
When we fit the result, we assume no $n$-$\pi$ ambiguity in each band. 
The ambiguity between bands is determined to minimize the sum of the square of the deviation.
We obtained $|{\rm RM_{obs}}|$ for the core region of J0928+4446 and J1018+0530 as $120\pm 7$~rad~m$^{-2}$ and $139\pm 5$~rad~m$^{-2}$, respectively.
Then, $|{\rm RM_{rest}}|$ is obtained as $1,010\pm 59$~rad~m$^{-2}$ and $1,200\pm 43$~rad~m$^{-2}$ for J0928+4446 and J1018+0530, respectively.
RM for J1159+0112 and J1405+4056 were not obtained because the detections of polarization only at one or two bands are inadequate to determine the $n$-$\pi$ ambiguity.

\subsection{Flux Variability}\label{sec:variability}

We examined the flux variability.
Between the FIRST survey and the NVSS,  we found no significant variability on the basis of significance of variability by defining $\Delta S = |S_1-S_2|$ and $\sigma_\mathrm{var}=(\sigma_1^{2}+\sigma_2^{2})^{1/2}$, where $S_i$ and $\sigma_i$ are total flux density and its uncertainty of $i$-th epoch data, respectively (Table~\ref{tbl:preflux}).
Errors were estimated to be 3\% in total flux density (Condon et al. 1998).
However, two-epoch observations are not enough to conclude that the sources are stable.
J1405+4056 shows $\Delta S > 3\sigma_{\rm var}$ at 5 GHz between the VIPS and our VLBA observation at 5~GHz;
however, we cannot rule out possible effects due to a different $uv$-coverage at the short baselines.
Thus, our verification of radio flux variability remains inconclusive. 

\section{DISCUSSION}\label{sec:discussion}

\subsection{Viewing Angle and Advancing Speed of Jets}\label{sec:viewingangle}

\begin{figure}
\includegraphics[width=0.5\textwidth]{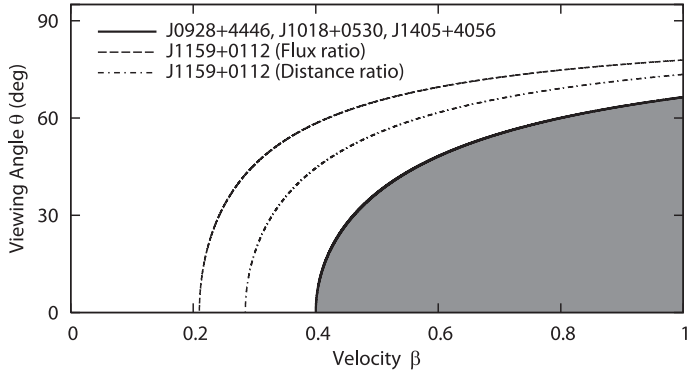}
\figcaption{
	Constraints on viewing angle $\theta$ and bulk speed $\beta$ of jets.  
	In the case of J1159+0112, results derived from flux density ratio (dashed) and core-jet distance ratio (dot-dashed) are shown.
	The upper limit of $\theta$ is set for the sources which show one-sided jet structure; J0928+4446, J1018+0530, and J1405+4056 (bold).
	Then, the filled area is allowed as properties of the sources with one-sided jet structure.	
\label{fig:betacos}
}
\end{figure}

Assuming an intrinsic symmetry of the jets, the apparent asymmetry of radio morphology with respect to the central engine results from a Doppler beaming effect.
The ratio of flux densities of an approaching to the receding components, $R_F$, is related to intrinsic jet velocity, $\beta$, and viewing angle, $\theta$, by $R_F = [(1+\beta \cos\theta) / (1-\beta\cos\theta) ]^{3-\alpha}$ \citep{1979ApJ...232...34B,1995PASP..107..803U}.
In the cases that counter jets were not detected, we apply $3\sigma$ noise upper limits.
We obtain $\beta\cos\theta>0.4$ for J0928+4446, J1018+0530, and J1405+4056, while $\beta\cos\theta\sim 0.2$ for J1159+0112.
As a result, the constraints on $\beta$ and $\theta$ are $0.4 <\beta <1$ and $\theta <66$~deg for J0928+4446, J1018+0530, and J1405+4056, while $0.2 <\beta <1$ and $\theta <77$~deg for J1159+0112 (Figure~\ref{fig:betacos}).

Alternatively, we can also obtain $\beta\cos\theta$ based on a core--jet distance ratio of an approaching to the receding components, given by $R_D = (1+\beta \cos\theta)/(1-\beta\cos\theta)$, which is applied to J1159+0112 with a two-sided structure.  
The apparent separation from the core (the component A) to a putative approaching component (the component D) is $\sim$90~mas and to the receding jet (the component E) is $\sim$50~mas.
Then, we obtain $\beta\cos\theta\sim 0.3$ (see Figure~\ref{fig:betacos}), which is nearly consistent with the result derived by the flux density ratio.
We obtain $0.3 <\beta <1$ and $\theta <73$~deg for J1159+0112 (Figure~\ref{fig:betacos}).

In summary, we gave moderate constraints of $\beta>0.4$ for J0928+4446, J1018+0530, and J1405+4056 while of  $\beta>0.3$ for J1159+0112 (Figure~\ref{fig:betacos}).
This indicates that two kinds of outflows are present in BAL quasars in terms of the speed \citep[cf.][]{2009PASJ...61.1389D}; one is the relatively fast nonthermal jet and the other one is slower wind ($\sim$0.1$c$).
The model of an accretion disk generating both radiation-force-driven wind \citep[e.g.,][]{2000ApJ...543..686P} and non-thermal jets with higher speeds \citep[e.g.,][]{2011ApJ...736....2O} should be applied to explain radio-loud BAL quasars.  
On the other hand, our estimations do not set so tight constraints on the orientation.

\subsection{Classification of Radio Sources}\label{sec:classification}

It is crucial to distinguish between blazars as pole-on viewed AGN and GPS radio sources as young radio sources for testing the orientation scheme and the evolution scheme of BAL quasars.    
Only on the basis of a spectral shape in a single-epoch observation, the flaring state of blazars could be misidentified as young radio sources.
Blazars show 
(i) long-term variability,
(ii) one-sided jet structure on mas scale, and
(iii) high degree of polarization.
In contrast, young radio sources show
(i) no significant spectral variability,
(ii) usually two-sided structure on mas scale, and
(iii) unpolarized radio emission in the core region at least at low frequency.
These criteria allow us to distinguish young radio sources from blazars in a convincing way \citep{2005A&A...432...31T,2007A&A...475..813O}.
In terms of the morphology (Section~\ref{sec:Imap}) and the polarization (Section~\ref{sec:pol}), J0928+4446, J1018+0530, and J1405+4056 can be classified as blazar candidates, while J1159+0112 as a young radio source.
In contrast, discrimination between blazars and young radio sources was inconclusive in terms of radio flux variability (Section~\ref{sec:variability}).
As a result, we found both blazar-type and young-radio-source-type BAL quasars in our targets which show flat or inverted spectra in \cite{2009PASJ...61.1389D}.

We note that all of our samples are originally selected only by absorption index (AI; \citealt{2002ApJS..141..267H,2006ApJS..165....1T}) but not by balnicity index (BI; \citealt{1991ApJ...373...23W}) which is a stricter criterion (see Table~\ref{tbl:sample}).
\cite{2008ApJ...687..859S} suggested that highest radio luminosities are preferentially found in AI-based but not BI-based BAL quasars and then a relation between luminosity distribution and Doppler beaming effect was discussed there (see also \citealt{2008MNRAS.386.1426K}, \citealt{2011arXiv1106.5916G}).
The finding of three blazar-type BAL quasars among the four could result from our radio-flux-limited selection.

\subsection{Constraints on the AGN wind}\label{sec:AGNwind}

\subsubsection{Geometry}\label{sec:windgeometry}

\begin{figure}
\includegraphics[width=0.5\textwidth]{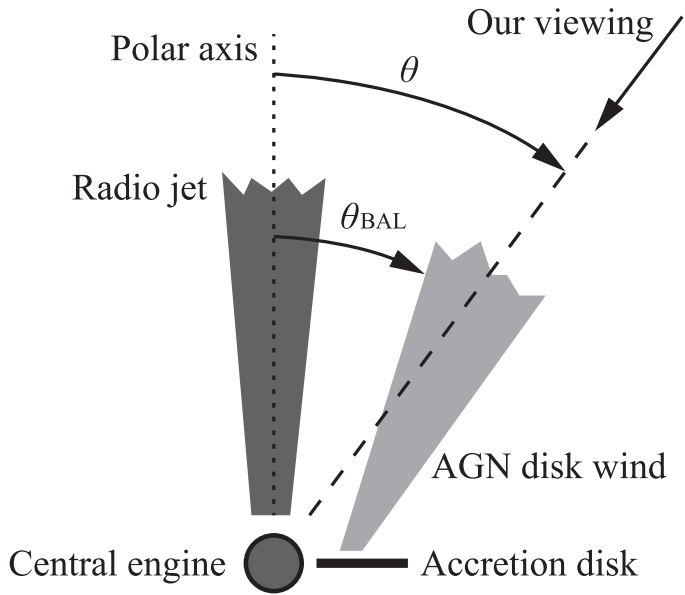}
\figcaption{
	Schematic picture of geometric relationship between wind and jet on the basis of \citet{2000ApJ...545...63E}.
	If we give constraint to our viewing, $\theta$, a restriction is also imposed on inclination angle of the wind, $\theta_{\rm BAL}$.
	The scale size of each component in the picture is arbitrary.
\label{fig:schematic}}
\end{figure}

The geometry of the AGN wind can be inferred from the viewing angle of the radio jet, which should be perpendicular to the innermost region of the accretion disk.
The AGN wind cuts across the line of sight to the central engine and the pc-scale non-thermal jets (Figure~\ref{fig:schematic}).  
The lower end of the range of opening angle for the AGN wind, $\theta_{\rm BAL}$, should be less than the upper limit of the viewing angle, $\theta$.
This constraint will be an intriguing comparison with the theoretical models of accretion disk because the AGN wind is thought to be lifted upward and accelerated to nearly edge-on from the disk by radiation force \citep{2000ApJ...543..686P}.
A radio imaging study is one of the most direct ways to test the orientation scheme.

We have set a constraint on $\beta\cos\theta$ for our target sources, using the flux density ratio of an approaching to the receding components.
We have also obtained $\beta\cos\theta$ using a core-jet distance ratio for J1159+0112 (Section~\ref{sec:viewingangle}).
The results are $\theta_{\rm BAL}<66$~deg for J0928+4446, J1018+0530, and J1405+4056, while $\theta_{\rm BAL}<73$~deg for J1159+0112 (Figure~\ref{fig:betacos}).  
These estimations give only mild constraints.  
According to the model presented by \cite{2000ApJ...545...63E}, the AGN wind bends outward to an opening angle of 60~deg with a divergence of 6~deg to give a covering factor of $\sim$10\%.
Although our finding of blazar-type BAL quasars strongly indicates pole-on-viewed AGN,
the derived inclinations are not strong constrains on the orientation of the AGN wind in the framework of the orientation scheme.
Stronger constraints can be obtained in a future observation because the capability of this method depends on an image dynamic range.

\subsubsection{Column Density}\label{sec:fromRM}
RM is related to physical properties through our line of sight as  
\begin{eqnarray}
\Bigl( \frac{\rm RM }{{\rm 1~m^{-2}~rad }} \Bigl)  
&=& 25 \Bigl(\frac{B_{\parallel}}{{\rm 1~mG}}\Bigl)  \Bigl(\frac{N}{{\rm 10^{20}~cm}^{-2}}\Bigl), \label{eq:RM2}
\end{eqnarray}
where $B_{\parallel}$ and $N$ are strength of averaged magnetic field parallel to the line of sight and column density of  thermal plasma, respectively.
Since RM is an integral quantity toward polarized sources, the observed RM comprises the contribution from the AGN wind, foreground medium (mainly Galactic contribution), and magnetized plasma associated with a non-thermal jet (e.g., \citealt{2002PASJ...54L..39A}).
RM due to the foreground medium contributes $\sim$30~rad~m$^{-2}$ in case of 45--70~deg in galactic latitude where our target sources range \citep{2001ARep...45..667P}.
As a result, we estimate RM due to the sources at $|{\rm RM_{rest}}|\sim 1010\pm 260$~rad~m$^{-2}$ and $|{\rm RM_{rest}}|\sim 1200\pm 260$~rad~m$^{-2}$ for J0928+4446 and J1018+0530, respectively,
These RM are within the range of the value for other radio sources \citep[e.g.,][]{1995PASJ...47..725I,2003ApJ...589..126Z}.
Although we cannot set constraints on the density of the AGN wind because the RM contains the contribution from the plasma associated with the jet,
further studies with multi-frequency radio polarimetric observations can provide the statistical estimates.

\subsection{Interpretations of the Radio Morphology of J1159+0112}\label{sec:discussionJ1159+0112}

Among four sources presented in this paper, it is notable that J1159+0012 shows multiple components with the extension  of more than 100~mas, which corresponds to the projected size of $\sim$1~kpc.
The bright central component with GHz-peaked spectrum can be interpreted as the core as often seen in the radio-loud quasars \citep{1979ApJ...232...34B}.
However, this source apparently exhibits the emissions in both sides of the core while most of quasars show an one-side jet  structure in VLBI-resolution scale.
The components B--D are relatively brighter than component E, suggesting that these components are the approaching jet components and the components E is the receding jet component.
The detection of the counter jet implies that the relativistic beaming is not significant at least in the component E.
The most likely explanation is that the counter jet is decelerated at the component E as a result of jet termination.
Strong polarized emission ($11.4\pm 1.7~$\%) is seen at the component D, and this polarized flux density constitutes the most of polarized flux density detected by VLA.
Multi-frequency VLA and single dish observations derived the intrinsic EVPA of $24\pm3$~deg which was taken into account the Faraday rotation \citep{2008MNRAS.388.1853M}.
The resultant magnetic field direction is $114\pm3$~deg, which is nearly perpendicular to the position angle of approaching jet.
Since the polarized flux density detected by our VLBA observations is almost equal to that by VLA, this magnetic field direction represents the one at the component D.
Both strong polarized flux density and magnetic field perpendicular to the jet are consistent as if this feature results from the compression of random magnetic field \citep{1980MNRAS.193..439L}, such as by shock.
These evidences allow us to infer that the components D and E are hot spots produced by the jet termination by ISM \citep[e.g.,][]{1981AJ.....86..833D,1982MNRAS.200..377T}.
It is noted that the most of radio-loud quasars show two-sided structure 
at low frequencies \citep{1994AJ....108..766B}.
Thus, the detection of the counter jet component in J1159+0012 is not surprising.
The absence of counterpart of the components B and C is also naturally explained if both approaching and receding jets are still relativistic before reaching to the hot spots.
One may think the lack of polarized emission from the component E is against above scenario but, if the component E possesses the similar level of fractional polarization with the component D, the polarized flux density is only $\sim$1~mJy, which is too diffuse to be detected by our observation.
Besides, the counter jet component could be affected by rather significant Faraday depolarization.
We need polarization observation with higher sensitivity particularly at higher frequency to confirm the polarized emission from the component E.

This source shows point-like structure in arcsecond scale (e.g., FIRST survey; \citealt{1995ApJ...450..559B}).
Two-sided structure with an angular size of about 200 mas, the same structure revealed by our observation, is seen on the image at 327~MHz by VLBA \citep{2009MNRAS.394L..61K}, whose total flux density is comparable to that measured by the Texas survey  \citep{1996AJ....111.1945D} at 365~MHz.
Therefore, the most of radio emission originates in the structure between the components D and E.
Even if we assume a small viewing angle as usually inferred for the quasars, for instance 10 degree, total extent of these radio emissions is $\sim$5~kpc.
This source size is relatively compact as compared to the classical double radio galaxies.
If we adopt typical hotspot velocity in young radio sources \citep{2002NewAR..46..263C}, the kinematic age of this source is $\sim$$10^4$--$10^5$~year. 
The source might be the product of recent episode of jet activity.

Another possible explanation for the core-dominated morphology is the change in Doppler factor and/or intrinsic jet power.  One-sided structure represented by the components A1 and A2 both at 5 and 8~GHz (see Figure~\ref{fig:J1159_I}) in the core and two-sided morphology represented by the components D and E in larger scale can be interpreted by the change in the jet angle to the line of sight.
Such a change in jet angle can be attributed to the precession of the jet axis.
One relevant observation is the BAL quasar J1048+3457.
This source shows a distorted morphology, which is produced by the jet precession \citep{2010ApJ...718.1345K}.
In contrast, even if the Doppler factor is constant with time, the brightness of the core (the component A) can also be explained by an increase in intrinsic luminosity of jets.
The newly ejected component in the core represents that the source might now be in an active phase with an increase in the jet power.
In this case, to form the much brighter core as compared to the extended emission, the activity should be intermittent. 
The activity of the core might cease after the components B--E were ejected.
Then, the time scale of the intermittency should be longer than the decay time of the extended components B--E \citep[e.g.,][]{2013arXiv1301.4759D}.
However, the farthest (or the oldest) component is the brightest, and thus some ad hoc ingredients are needed to reconcile this with the intermittency.

\section{CONCLUSION}\label{sec:conc}

Our VLBA polarimetric imaging for four radio-loud BAL quasars at 1.7, 5, and 8~GHz revealed the pc-scale radio structures of their non-thermal jets.  
J0928+446, J1018+0530, and J1405+4056 show  one-sided structures  in pc scales and polarized emissions in their cores.
Although radio flux variability is not confirmed in two-epoch observations, these characteristics are consistent with those of blazars.
These three sources are presumably pole-on-viewed AGNs, although our observations with limited image qualities provided only mild constraints on viewing angles that are not sufficient to compete with the orientation scheme.
On the other hand, J1159+0112 exhibits a two-sided structure across $\sim$1~kpc and no significant polarization in its central component, which shows a inverted spectrum.  
These characteristics are consistent with those of young radio sources.
The radio spectrum in the integrated flux density can be represented by a hybrid of a MHz-peaked spectrum component and a GHz-peaked spectrum component.
There are still several possible explanations for the GHz-peaked component and thus further study  (e.g., testing the variability) is needed to uncover the nature of the source.

\acknowledgments
We would like to acknowledge to G. Bruni, K.-H. Mack, F. M. Montenegro-Montes, K.~Ohsuga, and K.~Aoki for helpful discussions.
Kind advice concerning calibration of polarization with the EVLA data was given by J.~Linford.  
VLBA is a facility of NRAO, operated by Associated University Inc.~under cooperative agreement with the National Science Foundation (NSF).
This work made use of the Swinburne University of Technology software correlator, developed as part of the Australian Major National Research Facilities Programme and operated under license.
This research has made use of the VLA polarization monitoring program and the data from the UMRAO, which has been supported by the University of Michigan and by a series of grants from the NSF.
We also made use of the NASA/IPAC Extragalactic Database (NED), which is operated by the Jet Propulsion Laboratory, California Institute of Technology.
This work was partially supported by a Grant-in-Aid for Scientific Research (C; 21540250, A.D.), Global COE Program "the Physical Sciences Frontier" from the Japanese Ministry of Education, Culture, Sports, Science and Technology, and the Center for the Promotion of Integrated Sciences (CPIS) of Sokendai.


\clearpage

\clearpage

\clearpage

\clearpage

\clearpage

\clearpage

\clearpage

\clearpage

\clearpage

\clearpage
\begin{deluxetable}{ccccccccccc}
\tabletypesize{\scriptsize}
\tablecaption{Radio-loud BAL quasar sample for VLBA observation. \label{tbl:sample}}
\tablewidth{0pt}
\tablehead{\colhead{Object}	&	\colhead{SDSS name}	&	\colhead{$z$}	&	\colhead{BAL}	&	\colhead{AI}	&	\colhead{BI}	&	\colhead{$I^{\rm FIRST}_{\rm 1.4GHz}$}	&	\colhead{$S^{\rm FIRST}_{\rm 1.4GHz}$}	&	\colhead{$\log R_\ast$}	&	\colhead{$L^{\rm FIRST}_{\rm 1.4GHz}$}	&	\colhead{Ref.}	\\
\colhead{	}&\colhead{		}&\colhead{		}&\colhead{type}	&	\colhead{(km~s$^{-1}$)}	&	\colhead{(km~s$^{-1}$)}	&	\colhead{(mJy~beam$^{-1}$)}&	\colhead{(mJy)}	&	\colhead{	}&	\colhead{(erg~s$^{-1}$Hz$^{-1}$)}	&	\colhead{	}\\
\colhead{(1)}	&	\colhead{(2)}	&	\colhead{(3)}	&	\colhead{(4)}	&	\colhead{(5)}	&	\colhead{(6)}	&	\colhead{(7)}	&	\colhead{(8)}	&	\colhead{(9)}	&	\colhead{(10)}	&	\colhead{(11)}	}
\startdata
J0928+4446	&	J092824.13$+$444604.7	&	1.904 	&	Hi	&	293	&	0	&	156	&	162	&	2.8	&	3.7$\times 10^{33}$	&	a,b	\\
J1018+0530	&	J101827.85$+$053030.0	&	1.938 	&	Hi	&	441	&	0	&	284	&	297	&	3.3	&	6.8$\times 10^{33}$	&	a	\\
J1159+0112	&	J115944.82$+$011206.9	&	2.000 	&	Hi	&	2887	&	0	&	267	&	268	&	2.6	&	6.4$\times 10^{33}$	&	a,b,c,d	\\
J1405+4056	&	J140507.80$+$405657.8	&	1.993 	&	Hi	&	780	&	0	&	206	&	214	&	3.3	&	5.1$\times 10^{33}$	&	a
\enddata
\tablecomments{Column 1: object name in this paper.  Column 2: object name in the SDSS DR3. Column 3: redshift. Column 4: BAL classification. "Hi" denotes a HiBAL quasar.  Column 5: absorption index. Column 6: balnicity index.  Reference for Columns~2--6 is \cite{2006ApJS..165....1T}.  Column 7--8: FIRST 1.4-GHz peak flux density and integrated flux density \citep{1995ApJ...450..559B}.  Column 9: radio loudness \citep{2009PASJ...61.1389D}.  Column 10: 1.4-GHz radio luminosity calculated by equation 1 in \cite{2008ApJ...687..859S} assuming spectral index of $\alpha =-0.7$.  Column 11: references where the BAL quasar appears in the literature.}

\tablerefs{a: \citealt{2006ApJS..165....1T}, b: \citealt{2009ApJ...692..758G}, c: \citealt{2009MNRAS.399.2231S}, d: \citealt{1984ApJ...287..549B}}																					
\end{deluxetable}

\clearpage															
\begin{deluxetable}{cccccc}
\tabletypesize{\scriptsize}
\tablecaption{Flux densities and spectral index of each component. \label{tbl:flux}}
\tablewidth{0pt}
\tablehead{\colhead{Object}	&	\colhead{Com-}	&	\multicolumn{3}{c}{Flux density}			&	\colhead{Spectral index}	\\
\colhead{	}&	\colhead{ponent}	&	\colhead{1.7~GHz}	&	\colhead{4.9~GHz} &	\colhead{8.3~GHz}			&		\colhead{	}	\\	
\colhead{	}&	\colhead{	}&	\colhead{(mJy)}	&	\colhead{(mJy)}	&	\colhead{(mJy)}	&\colhead{ }}
\startdata
J0928+4446	&	A	&	164.9 	$\pm$	8.2 	&	206.7 	$\pm$	10.3 	&	258.3 	$\pm$	12.9 	&	0.26 	$\pm$	0.05 	\\	
	&	B	&	14.8 	$\pm$	0.7 	&	5.6 	$\pm$	0.3 	&	3.1 	$\pm$	0.2 	&	$-$0.96 	$\pm$	0.04 	\\	
	&	C	&	6.1 	$\pm$	0.3 	&	0.7 	$\pm$	0.0 	&		$<$	0.6 	&	$-$2.01 	$\pm$	0.09 	\\	
	&	D	&	1.1 	$\pm$	0.1 	&		$<$	0.8 	&		$<$	0.6 	&		\nodata		\\
J1018+0530	&	Total	&	357.7 	$\pm$	17.9 	&	368.4 	$\pm$	18.4 	&	395.1 	$\pm$	19.8 	&	0.06 	$\pm$	0.03 	\\
J1159+0112	&	A	&	41.9 	$\pm$	2.1 	&	99.9 	$\pm$	5.0 	&	135.6 	$\pm$	6.8 	&	0.74 	$\pm$	0.06 	\\	
	&	B	&	18.9 	$\pm$	0.9 	&	2.1 	$\pm$	0.1 	&		$<$	0.8 	&	$-$2.07 	$\pm$	0.09 	\\	
	&	C	&	29.1 	$\pm$	1.5 	&	7.5 	$\pm$	0.4 	&		$<$	0.8 	&	$-$1.27 	$\pm$	0.09 	\\	
	&	D	&	66.9 	$\pm$	3.3 	&	16.2 	$\pm$	0.8 	&		$<$	0.8 	&	$-$1.33 	$\pm$	0.09 	\\	
	&	E	&	11.9 	$\pm$	0.6 	&	0.9 	$\pm$	0.0 	&		$<$	0.8 	&	$-$2.47 	$\pm$	0.09 	\\
J1405+4056	&	A	&	222.7 	$\pm$	11.1 	&	227.1 	$\pm$	11.4 	&	214.2 	$\pm$	10.7 	&	$-$0.02 	$\pm$	0.03 	\\	
	&	B	&	18.8 	$\pm$	0.9 	&	11.7 	$\pm$	0.6 	&	1.5 	$\pm$	0.1 	&	$-$1.49 	$\pm$	0.58 	\\	
	&	C	&	16.4 	$\pm$	0.8 	&	5.3 	$\pm$	0.3 	&	3.1 	$\pm$	0.2 	&	$-$1.04 	$\pm$	0.02 
\enddata
\end{deluxetable}

\clearpage																											
\begin{deluxetable}{cccccccc}																											
\tabletypesize{\scriptsize}																											
\tablecaption{Polarized flux densities and degree of polarization of each component. \label{tbl:pol_flux}}																											
\tablewidth{0pt}																											
\tablehead{																											
Object	&	Com-	&	\multicolumn{3}{c}{Polarized Flux density}											&	\multicolumn{3}{c}{Degree of polarization}											\\
	&	ponent	&	1.7~GHz			&	4.9~GHz			&	8.3~GHz			&	1.7~GHz			&	4.9~GHz			&	8.3~GHz			\\
	&		&	(mJy)			&	(mJy)			&	(mJy)			&	(\%)			&	(\%)			&	(\%)			
}																											
\startdata																											
J0928+4446	&	A	&	7.8 	$\pm$	0.8 	&	8.6 	$\pm$	0.9 	&	9.5 	$\pm$	1.0 	&	4.6 	$\pm$	0.7 	&	4.6 	$\pm$	0.7 	&	4.4 	$\pm$	0.7 	\\
J1018+0530	&	Total	&	13.8 	$\pm$	1.4 	&	7.8 	$\pm$	0.8 	&	6.9 	$\pm$	0.7 	&	3.9 	$\pm$	0.6 	&	2.2 	$\pm$	0.3 	&	1.8 	$\pm$	0.3 	\\
J1159+0112	&	D	&	7.6 	$\pm$	0.8 	&		$<$	1.2 	&		$<$	1.1 	&	11.4 	$\pm$	1.7 	&		$<$	7.4 	&		\nodata		\\
J1405+4056	&	A	&		$<$	0.9 	&	2.9 	$\pm$	0.3 	&	4.7 	$\pm$	0.5 	&		$<$	0.3 	&	1.3 	$\pm$	0.2 	&	2.3 	$\pm$	0.4 	\\
	&	C	&	2.3 	$\pm$	0.3\tablenotemark{a}	&		$<$	0.9 	&		$<$	0.9 	&	13.7 	$\pm$	2.7\tablenotemark{a} 	&		$<$	17.0 	&		$<$	29.0 	
\enddata																											
\tablenotetext{a}{Only detected at 1.663~GHz.}																											
\end{deluxetable}

\clearpage																																			
\begin{deluxetable}{cccccccccc}																																			
\tabletypesize{\scriptsize}																																			
\tablecaption{Electric vector position angle and Faraday rotation measure of each component \label{tbl:EVPA}}																																			
\tablewidth{0pt}																																			
\tablehead{																																			
Object	&	Com-	&	\multicolumn{6}{c}{Electric vector position angle}																							&	Observed			&	Rest-frame			\\
	&	ponent	&	1.663~GHz			&	1.671~GHz			&	4.644~GHz			&	5.095~GHz			&	8.111~GHz			&	8.580~GHz			&		RM		&		RM		\\
	&		&	(deg)			&	(deg)			&	(deg)			&	(deg)			&	(deg)			&	(deg)			&	(rad m$^{-2}$)			&	(rad m$^{-2}$)			
}\startdata																																			
J0928+4446	&	A	&	234 	$\pm$	9 	&	237 	$\pm$	9 	&	30 	$\pm$	3 	&	39 	$\pm$	3 	&	22 	$\pm$	2 	&	21 	$\pm$	2 	&	120 	$\pm$	7 	&	1012 	$\pm$	59 	\\
J1018+0530	&	Total	&	339 	$\pm$	9 	&	340 	$\pm$	9 	&	107 	$\pm$	3 	&	111 	$\pm$	3 	&	96 	$\pm$	3 	&	87 	$\pm$	3 	&	139 	$\pm$	5 	&	1200 	$\pm$	43 	\\
J1159+0112	&	D	&	152 	$\pm$	9 	&	159 	$\pm$	9 	&		\nodata		&		\nodata		&		\nodata		&		\nodata		&		\nodata		&		\nodata		\\
J1405+4056	&	A	&		\nodata		&		\nodata		&	101 	$\pm$	4 	&	114 	$\pm$	4 	&	91 	$\pm$	5 	&	94 	$\pm$	4 	&		\nodata		&		\nodata		\\
	&	C	&	14 	$\pm$	10 	&		\nodata		&		\nodata		&		\nodata		&		\nodata		&		\nodata		&		\nodata		&		\nodata		
\enddata																																			
\end{deluxetable}

\clearpage
\begin{deluxetable}{ccccccc}
\tabletypesize{\scriptsize}
\tablecaption{Total flux densities and significance of variability.\label{tbl:preflux}}
\tablewidth{0pt}
\tablehead{\colhead{Object}	&	\colhead{NVSS}			&	\colhead{FIRST}			&	\colhead{$\Delta S/\sigma_{\rm var}$} 	&	\colhead{VIPS}			&	\colhead{our VLBA}			&	\colhead{$\Delta S/\sigma_{\rm var}$} 	\\
\colhead{	}&	\colhead{1.4~GHz}			&	\colhead{1.4~GHz}			&	\colhead{1.4~GHz} 	&	\colhead{5~GHz}			&	\colhead{5~GHz}			&	\colhead{5~GHz}	\\
\colhead{	}&	\colhead{(mJy)}			&	\colhead{(mJy)}			&	\colhead{	}&	\colhead{(mJy)}			&	\colhead{(mJy)}			&\colhead{ }}
\startdata																			
J0928+4446	&	170.2 	$\pm$	5.1 	&	162.2 	$\pm$	4.9 	&	1.1 	&	251.9 	$\pm$	12.6 	&	213.1 	$\pm$	10.7 	&	2.3 	\\
J1018+0530	&	277.8 	$\pm$	8.3 	&	296.6 	$\pm$	8.9 	&	1.6 	&		\nodata		&	368.4 	$\pm$	18.4 	&	\nodata	\\
J1159+0112	&	275.6 	$\pm$	8.3 	&	268.5 	$\pm$	8.1 	&	0.6 	&		\nodata		&	126.6 	$\pm$	6.3 	&	\nodata	\\
J1405+4056	&	205.9 	$\pm$	6.2 	&	214.0 	$\pm$	6.4 	&	0.9 	&	193.4 	$\pm$	9.7 	&	244.1 	$\pm$	12.2 	&	3.3 	
\enddata																					
\end{deluxetable}

\end{document}